\documentclass[epj]{svjour}

\usepackage{epsfig,psfrag}
\usepackage{amssymb}

\begin{document}

\title{Zero bias anomaly in the density of states of low-dimensional metals}
\author{Lorenz Bartosch and Peter Kopietz}

\institute{Institut f\"ur Theoretische Physik, Universit\"at Frankfurt, Robert-Mayer-Str. 8-10, D-60054 Frankfurt am Main, Germany}
\date{August 20, 2001}

\abstract{
We consider the effect of Coulomb interactions
on the average density of states (DOS)
of disordered low-dimensional metals  
for temperatures $T$ and frequencies $\omega$ smaller than the
inverse elastic life-time $1/ \tau_0$.
Using the fact that long-range Coulomb interactions 
in two dimensions ($2d$) generate
$\ln^2$-singularities in the DOS
$\nu ( \omega  )$ 
but only  $\ln$-singularities in the conductivity 
$\sigma ( \omega) $,  we can re-sum the most singular contributions to the
average DOS via a simple gauge-transformation.
If $\lim_{ \omega \rightarrow 0 } \sigma ( \omega )  > 0$,
then  a {\it{metallic Coulomb gap}} 
$\nu ( \omega ) \propto  | \omega | / e^4$ 
appears in the DOS at $T=0$ for frequencies below a certain crossover
frequency $\Omega_2$ which depends on the
value of the DC conductivity $\sigma ( 0 )$.
Here, $-e$ is the charge of the electron.
Naively adopting the same procedure to calculate the DOS in quasi $1d$
metals, we find 
$\nu(\omega) \propto (|\omega|/\Omega_1)^{1/2} \exp(-\Omega_1/|\omega|)$ at $T = 0$, 
where $\Omega_1$ is some interaction-dependent 
frequency  scale.  However, we argue that in quasi $1d$ 
the above gauge-transformation method  is on less firm grounds
than in $2d$.
We also discuss the behavior of 
the DOS at finite temperatures and give numerical results for the
expected  tunneling conductance that can be compared with 
experiments.
}
       
\maketitle


%
%
\section{Introduction}
\setcounter{section}{1}
\label{sec:intro}
In the early eighties, Altshuler and Aronov \cite{Altshuler85} 
perturbatively studied the effect of electron-electron interactions 
on the density of states (DOS) of low-dimensional weakly disordered 
interacting electronic systems.
For temperatures $T$ and frequencies $\omega $ 
smaller than the inverse elastic life-time $1 / \tau_0$, they found that
in reduced dimensions
the interplay between disorder and electron-electron interactions 
gives rise to singular corrections to the averaged DOS. 
In two dimensions ($2d$), the long-range Coulomb interaction gives rise to a 
$\ln^2$-correction to the average DOS \cite{Altshuler85},
 \begin{equation}
  \label{eq:Altshuler2}
  \nu(\omega) \sim \nu_0 \left[ 1 -  \frac{r_0}{4} \ln ( | \omega | \tau_0 ) \ln ( | \omega | \tau_1 ) \right] 
 \;  \; , \; \; 2d
 \; ,
\end{equation}
where
 \begin{equation}
 r_0 = \frac{ 1 }{ ( 2 \pi )^2 \nu_0 D_0 } = \frac{ 2 e^2}{ (2 \pi )^2 \sigma_0} = \frac{1}{\pi k_F \ell }
 \label{eq:r0def}
 \end{equation}
is a dimensionless measure for the resistance of the system at frequency scale
$\omega \approx \tau_0^{-1}$ (where $\sigma_0$ is the Drude conductivity), 
 and the interaction-dependent time $\tau_1$ is given by
 \begin{equation}
 \tau_1 = \frac{1}{ D_0^2 \kappa^4 \tau_0 } = \frac{4 \tau_0}{ ( \kappa \ell )^4 }
 \; .
 \end{equation}
Here, $\kappa = 2 m e^2$ is the Thomas-Fermi screening wave-vector in two dimensions, 
$\nu_0 = m / 2 \pi$ is the
DOS at the Fermi energy (per spin projection) of electrons with effective mass $m$ and charge $-e$,
$D_0 = v_F \ell /2$ is the diffusion coefficient in $2d$, and $\ell = v_F \tau_0$
is the elastic mean free path.
We use units such that $\hbar = k_B =1$.
Note that for a good metal at high densities the Thomas-Fermi screening length is short compared with  the
elastic mean free path ($\kappa \ell \gg 1$) so that $ \tau_1 \ll \tau_0$.   
In 
quasi $1d$ metallic wires  (which consist of many transverse channels but 
permit diffusive motion only 
in one direction) the leading correction to the average DOS is \cite{Altshuler85}
\begin{equation}
  \label{eq:Altshuler1}
  \nu(\omega) \sim \nu_0 \left[ 1 - \sqrt{\frac{4 \Omega_{1}}{\pi |\omega|}}\ \right] 
 \;  \; , \; \;  \mbox{quasi $1d$}
\;,
\end{equation}
where $\nu_0 = 1/ ( \pi v_F) $  is the DOS per spin 
in $1d$, and the frequency scale $\Omega_1$ 
depends on 
the effective electron-electron interaction constant $f_0$,
\begin{equation}
  \label{eq:omega.ast}
  \Omega_{1} = \frac{f_0^2}{32 \pi D_0} \;.
\end{equation}
Here, $D_0 = v_F \ell $ is the (bare) diffusion coefficient in quasi $1d$.

Obviously, the correction terms 
in Eqs.\ (\ref{eq:Altshuler2}) and (\ref{eq:Altshuler1})  
diverge for $ \omega \rightarrow 0$,  so that at low frequencies these 
perturbative expressions cease to be valid.
What is the true low-frequency behavior of the DOS of disordered metals in 
reduced dimensions?  
The answer to this question is relevant 
for a number of recent tunneling experiments \cite{Bielejec01,Bachtold99,Bachtold00}, where
a strong suppression of the tunneling conductance
$G ( V )$ as a function of the applied voltage has been observed (zero bias anomaly).
The tunneling conductance is related to the DOS via
$G ( V ) \propto \nu ( \omega = e V )$, so that the experimentally
observed zero bias anomaly in the tunneling conductance reflects the
strong suppression of the average DOS at the Fermi energy.

In $2d$,
the low-temperature behavior of the DOS of a 
strongly correlated disordered metal 
has recently been measured by Bielejec {\em et al.} \cite{Bielejec01}.
While at higher temperatures they found a logarithmic correction, 
at the lowest temperatures they found a stronger-than linear suppression of the DOS, which 
has been interpreted in terms of a hard correlation gap. 
The knowledge of the  low-energy behavior of the DOS 
of a $2d$ disordered metal with a finite conductivity might also be important 
to gain a better 
understanding of  the physical mechanism
that is responsible for the 
apparent metal-insulator transition in doped semiconductor 
devices \cite{Kravchenko95,Abrahams01}.
An intensely studied quasi $1d$ system  
where under certain conditions the electrons
propagate diffusively in only one direction   
are  multi-wall Carbon nanotubes  \cite{Bachtold99,Bachtold00}.

Let us briefly review previous calculations of the zero bias anomaly.
In $2d$, the first attempt to determine the true low-frequency asymptotics of  
$\nu ( \omega )$ was apparently due to Finkelstein \cite{Finkelstein83} who found
that $ \nu ( \omega ) \propto | \omega |^{1/4}$ for $ \omega \rightarrow 0$. However, in 
the derivation of this 
result he assumed that the conductivity $\sigma ( \omega )$ diverges
logarithmically 
for $ \omega \rightarrow 0$. The behavior of the DOS of $2d$
disordered electrons with a finite
conductivity was not calculated by Finkelstein.
Later this problem was reconsidered by Belitz and Kirkpatrick \cite{Belitz93}, who found that 
for frequencies exceeding the crossover frequency
 \begin{equation} 
 \Omega_{2} \equiv \tau_0^{-1}  \exp[ - 1 / r_0 ] 
 \; ,
 \label{eq:omega2}
 \end{equation}
the perturbative expression given in Eq.\ (\ref{eq:Altshuler2}) can actually be
exponentiated, so that
 \begin{equation}
  \label{eq:Belitz}
  \nu(\omega) \approx \nu_0 \exp \left[ -  \frac{r_0}{4} \ln ( | \omega | \tau_0 ) \ln ( | \omega | \tau_1 ) \right] 
 \; .
\end{equation}
This expression has been re-derived
in different ways by several authors \cite{Levitov97,Kopietz98,Kamenev99}.
Note, however,  that Eq.\ (\ref{eq:Belitz})  
is not valid for
frequencies smaller than the crossover frequency $\Omega_{2}$ 
defined in Eq.\ (\ref{eq:omega2}) \cite{Belitz93,Kopietz98,Kamenev99}.
A simple interpolation formula for the DOS, which yields a physically
sensible result even for $\omega \rightarrow 0$,
has been proposed by one of us in Ref.\ \cite{Kopietz98}.
This formula is based on a re-summation of the
leading $\ln^2$-singularities to all orders in perturbation theory,
consistently neglecting sub-leading terms that involve only logarithmic corrections.
In this approximation, one obtains
 \begin{equation}
 \nu ( \omega ) \approx \nu_0 \frac{2}{\pi} \int_{\tau_0}^{\infty} d t 
 \frac{ \sin ( | \omega | t )}{ t } \exp \left[ - \frac{r_0}{4} \ln ( t / \tau_0 ) \ln ( t / \tau_1 ) \right]
 \; .
 \label{eq:Cbgap}
 \end{equation}
For $ | \omega | \gtrsim \Omega_{2} = \tau_0^{-1} \exp[ - 1 / r_0 ]$, 
this expression reduces to Eq.\ (\ref{eq:Belitz}).
Note that Eq.\ (\ref{eq:Cbgap}) amounts to an exponentiation of the perturbative result
in the time domain,
whereas  in Eq.\ (\ref{eq:Belitz}) 
the perturbation series is exponentiated in  frequency space.
In the limit $\omega \rightarrow 0$, the leading term in the asymptotic expansion of
Eq.\ (\ref{eq:Cbgap}) is
 \begin{equation}
 \nu ( \omega ) \sim \nu_0  \frac{ 4 }{  \pi^{1/2} }
 \left( \frac{ \tau_1}{ \tau_0 r_0 } \right)^{1/2} 
 \frac{ | \omega |   }{ \Omega_{2}  }
 \; .
 \label{eq:leadingasym}
 \end{equation}
Noting that  $\sqrt{ \tau_0 \tau_1 } = r_0 / ( \nu_0 e^4 )$, 
Eq.\ (\ref{eq:leadingasym}) can also be written as
 \begin{equation}
 \nu ( \omega ) \sim C \frac{| \omega | }{ e^4} 
 \; ,
 \label{eq:nuasym}
 \end{equation}
where the numerical constant $C$ is given by
 \begin{equation}
 C =  4 ( r_0 / \pi )^{1/2} \exp [  1 / r_0 ]  
 \; .
 \label{eq:Cres}
 \end{equation}
Surprisingly, Eq.\ (\ref{eq:nuasym}) resembles the well-known classical 
Efros-Shklovskii Coulomb gap of two-dimensional electrons in the localized regime, where the DC conductivity $\sigma(0)$ vanishes \cite{Efros85}.
Note, however, that in the derivation of Eq.\ (\ref{eq:nuasym}) we have assumed
that the DC conductivity $\sigma ( 0 )$ remains finite, i.e.  the electrons are assumed to be 
{\it{delocalized}}. 
The difference between the localized and the delocalized regime manifests itself in the dimensionless prefactor $C$: whereas
in the case of the classical Coulomb gap of localized electrons the constant
$C$ is a number of the order of unity that depends on the geometry of the underlying lattice,
for the quantum Coulomb gap discussed here,
$C$ depends on the dimensionless conductivity of the system. 
The existence of the Coulomb gap in the delocalized regime of a disordered interacting $2d$ metallic system was also found numerically by Efros and Pikus \cite{Efros95}. More recently, an intermediate delocalized phase in small clusters of disordered interacting electrons has been found numerically in Ref.\ \cite{Benenti99}.

In deriving Eq.\ (\ref{eq:Cres}),  we have implicitly assumed that the
DC conductivity $\sigma ( 0 )$ does not deviate significantly from the
conductivity $\sigma ( \omega )$ at frequency $ \omega \approx 1 / \tau_0$.
We have argued elsewhere \cite{Kopietz98} that a finite renormalization of the conductivity
can be approximately taken into account be replacing $ r_0 \rightarrow r_{\ast}$
in Eq.\ (\ref{eq:Cres}) where $r_{\ast} =  e^2 / ( 2 \pi^2 \sigma ( 0 ) )$ corresponds
to the  true  DC conductivity of the system. 
The connection between low-frequency behavior of the conductivity and the
DOS has also been emphasized by Nazarov \cite{Nazarov89}, and by
Levitov and Shytov \cite{Levitov97}.

The zero bias anomaly in $1d$ has so far received much less attention than the
corresponding anomaly in $2d$. Recent experiments on metallic Carbon 
nano\-tubes have motivated Mishchenko {\em et al.} \cite{Mishchenko01} to 
study the fate of the perturbative Altshuler-Aronov correction
in $1d$ at low frequencies.
They found that long-range Coulomb interactions 
in a quasi $1d$ metal lead
for $\omega \rightarrow 0$ to an exponentially small  DOS,
 \begin{equation}
 \nu ( \omega ) \propto \exp \left[ 
 - \frac{ \epsilon_0 }{ | \omega |} \ln \left( \frac{ \epsilon_1}{ | \omega |} \right) 
\right]
 \label{eq:numishchenko}
 \; ,
 \end{equation}
where $\epsilon_0$ and $\epsilon_1$ are some finite energy scales.
We shall further comment on this result below. 
For frequencies exceeding a crossover scale 
$ D_0/(2\pi R)^2$, where $R$ is the radius of the  
nano\-tube, Egger and Gogolin \cite{Egger01} 
found a crossover to two-dimensionality, 
which results in a power-law dependence of the DOS.
However, below this crossover scale, $1d$ behavior is expected.

The rest of this work is organized as follows: In Sec.\ \ref{sec:gauge} we 
critically examine the non-perturbative method leading to
the above results for the zero bias anomaly. In particular, we 
argue that only in the case of long-range Coulomb interactions
in $2d$ the method can be formally justified; in particular,
in quasi $1d$ the approximations leading to the result
given in Eq.\ (\ref{eq:numishchenko}) are uncontrolled,  so that
we have serious doubts whether 
the exponential suppression
of the DOS given in Eq.\ (\ref{eq:numishchenko}) is the physically correct low-frequency
behavior in quasi $1d$. 
In Sec.\ \ref{sec:2d}
we generalize
the calculation of Ref. \cite{Kopietz98}
to finite temperatures and present numerical results for the frequency-dependence
of the average DOS  which in principle can be  compared with 
experimentally measured tunneling conductances.
In Sec.\ \ref{sec:1d} we discuss the DOS in quasi $1d$ and explain 
why in this case the calculation cannot
be performed with the same confidence as in $2d$.
Finally, we end
with a brief summary.

\section{Summing the leading singularities via a gauge transformation}
\label{sec:gauge}

The average DOS of an interacting Fermi system at finite temperature $T=1/\beta$ can be written in terms of the disorder-averaged Green function at coinciding space points ${\mathcal{G}}(\omega) \equiv \overline {{\mathcal{G}}({\bf r},{\bf r},\omega)}$ as
\begin{equation}
  \label{eq:def:DOS}
  \nu(\omega,T) = - \frac{1}{\pi} \coth \left( \frac{\beta \omega}{2} \right) \textrm{Im}\, {\mathcal{G}}(\omega) \;.
\end{equation}
To make contact with the disorder-averaged imaginary-time Green function at coinciding space points $G(\tau)$, we notice that due to particle-hole symmetry near the Fermi energy we have
\begin{equation}
  \label{eq:Greenfunction}
  {\mathcal{G}}(\omega) = -2 \int_0^{\infty} dt\, \sin \omega t \ G(\tau \to it +0^{+}) \;.
\end{equation}
The problem of calculating the average DOS is now
reduced to the problem of calculating
the disorder-averaged imaginary-time Green function. 
An attempt to calculate $\nu ( \omega )$ within perturbation theory fails at low
frequencies because the perturbative expansion of $\nu ( \omega )$ is
plagued by singularities, see Eqs.\ (\ref{eq:Altshuler2}) and (\ref{eq:Altshuler1}). In $2d$ with long-range Coulomb interactions, these singularities diverge as $\ln^2 \omega$. 
At the same time, however, 
the perturbative expansion of the conductivity $ \sigma ( \omega )$ contains only
less severe $\ln \omega$-singularities \cite{Altshuler85,Finkelstein83}.
As a consequence, for sufficiently small frequencies, 
we may re-sum the most singular  terms in the expansion of the average 
DOS {\em without considering
simultaneously a similar re-summation for the average conductivity},
because we know a priori that $\ln^2$-singularities do not
appear  in the calculation of $\sigma ( \omega )$. In this sense,
the problem of calculating the average DOS decouples from the problem
of calculating the average conductivity. 
As shown in Ref.\ \cite{Kopietz98}, under these conditions
we may  sum the most singular contributions to the average DOS via a 
simple gauge transformation, which at finite temperature $T$ leads to the
following expression for the imaginary time Green function,
\begin{equation}
  \label{eq:G=G0*exp}
  G(\tau) \approx G_0(\tau) e^{Q(\tau)} \;,
\end{equation}
where $G_0(\tau)$ is the Green function of free fermions,
\begin{equation}
  \label{eq:G0}
  G_0(\tau)= - \nu_0 \, \frac{\pi/\beta}{\sin (\pi \tau /\beta)} \;,
\end{equation}
and the Debye-Waller factor $Q(\tau)$ is given by \cite{footnote1}
\begin{equation}
  \label{eq:DebyeWaller}
  Q(\tau) = -\frac{1}{\beta V} \sum_{{\bf q},\omega_m} \frac{f^{\rm  RPA}_{{\bf q},i\omega_m}}{(D_0 {\bf{q}}^2 + |\omega_m|)^2} 
\left[1-\cos ( \omega_m \tau ) \right] \;.
 \end{equation}
Here, $f^{\rm RPA}_{{\bf q},i\omega_m}$ is the dynamically screened average 
Coulomb interaction, which is a function of momentum ${\bf q}$ 
and bosonic Matsubara frequency $\omega_m = 2\pi m/\beta$.
The volume of the system is denoted by $V$.
It is important to emphasize that  in deriving Eqs.\ (\ref{eq:G=G0*exp}-\ref{eq:DebyeWaller})
the sub-leading logarithmic 
corrections have been ignored, so that it would be inconsistent to
retain  sub-leading terms involving only a single logarithm
in the evaluation of the Debye-Waller factor.  

 Apparently, the  above gauge-transformation method to 
re-sum the leading terms in the
perturbative expansion of the single-particle Green function 
has first been used by  Nazarov \cite{Nazarov89}.
Later, several  authors employed this technique  
to calculate the DOS of  disordered interacting 
electrons \cite{Levitov97,Kopietz98,Kamenev99}. 
However, one should keep in mind that in practice this method relies
on the fact that
the problem of calculating the average single-particle Green function 
decouples in the sense discussed above from the problem of calculating the conductivity.
In particular, for $2d$ disordered electrons subject to
{\em short-range interactions},
the perturbative calculation of 
$\nu ( \omega )$ and $\sigma ( \omega )$ both involve
 $\ln \omega $-singularities , so that 
a naive application of the above gauge-transformation trick 
for short-range interactions in $2d$ is
at least problematic. This is also the case for diffusive quasi $1d$ electrons, where
the perturbative calculation of both
$\nu ( \omega )$ and $\sigma ( \omega )$ leads to 
$| \omega |^{-1/2}$-singularities \cite{Altshuler85}. Hence, also in this case
it is problematic to  calculate the average DOS using
 Eqs.\ (\ref{eq:G=G0*exp}--\ref{eq:DebyeWaller}) 
without considering simultaneously the  low-frequency behavior 
of the conductivity. We shall come back to this point
in Sec.\ \ref{sec:1d}.

\section{Zero bias anomaly in 2d}
\label{sec:2d}
\begin{figure}[htb]
\begin{center}
\psfrag{omega}{$\omega/\Omega_2$}
\psfrag{dos}{$\nu(\omega )/\nu_0$}
\psfrag{0}{$0$}
\psfrag{0.2}{$0.2$}
\psfrag{0.25}{$0.25$}
\psfrag{-0.25}{$-0.25$}
\psfrag{0.4}{$0.4$}
\psfrag{0.5}{$0.5$}
\psfrag{-0.5}{$-0.5$}
\psfrag{0.6}{$0.6$}
\psfrag{0.8}{$0.8$}
\psfrag{1}{$1$}
\psfrag{-1}{$-1$}
\epsfxsize8.0cm
\epsfbox{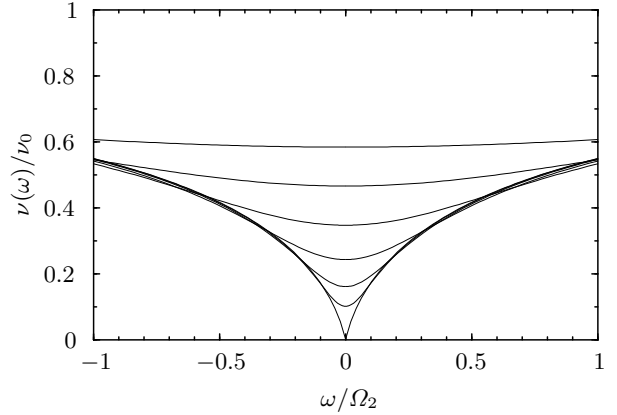}
\caption{Graph of the average DOS  $\nu(\omega)$ in $2d$ for various 
temperatures, see Eq.\ (\ref{eq:2d:DOS}).
We have chosen 
$r_0 \to r_{\ast} =1/\pi$ and $\tau_1 = \tau_0$. 
Curves from top to bottom are for $T/ \Omega_2 = 0.8,0.4,0.2,0.1,0.05,0.025,0$.}
\label{fig:2d:DOS}
\end{center}
\end{figure}

In $2d$ the RPA-interaction is given by \cite{Altshuler85}
\begin{equation}
  f^{\rm RPA}_{ {\bf{q}} , i \omega_m } \approx (2 {{D}}_0 \nu_0)^{-1} | \omega_m | / 
  {\bf{q}}^2 
 \label{eq:2d:RPA-interaction} \; .
\end{equation}
This expression is valid in the frequency-momentum regime
\begin{equation}
  | \omega_m | / {{D}}_0 \kappa \ll | {\bf{q}} | \ll
( | \omega_m | / {{D}}_0 )^{1/2} \; , \quad | \omega_m | \ll \tau_0^{-1}
 \label{eq:regime}
 \; ,
\end{equation}
which is responsible for the $\ln^2$-corrections to the DOS. 
Converting the sum over Matsubara frequencies into an integral, 
the Debye-Waller factor (analytically continued to real time) may be written as 
\begin{eqnarray}
  Q(it) & = & \frac{r_0}{2} \int_0^{\tau_0^{-1}} \frac{d\omega}{\omega} \,\ln\left(\frac{\omega}{D_0 \kappa^2} \right) \nonumber \\
  & & \qquad \times \left[\frac{1-\cos(\omega t)}{\tanh(\beta \omega/2)} + 
i \sin(\omega t) \right] \;,
  \label{eq:d2:DebyeWaller2}
\end{eqnarray}
where we have taken the thermodynamic limit $ V \rightarrow \infty$.
To be consistent with the approximation made in Eq.\ (\ref{eq:G=G0*exp}), 
we retain only $\ln^2$-singularities
and ignore all terms involving only single logarithms: Within this approximation the 
imaginary part of $Q(it)$ vanishes, and for arbitrary temperatures we find $Q (it) \sim -\frac{r_0}{4} \ln(t/\tau_1) \ln(t/\tau_0)$.
The DOS at finite temperature $T=1/\beta$ is now given by
\begin{eqnarray}
  \label{eq:2d:DOS}
    \nu(\omega , T ) & \approx & \nu_0 \coth \left( \frac{\beta \omega}{2} \right) \frac{2}{\beta} \int_{\tau_0}^{\infty} dt \, \frac{\sin (\omega t)}{\sinh(\pi t/ \beta)} \qquad \nonumber \\
    & & \qquad \quad \times 
\exp \left[ -\frac{r_0}{4} \ln(t/\tau_1) \ln(t/\tau_0) \right] \;.
\end{eqnarray}
In the limit  $T \rightarrow 0$ this equation reduces to Eq.\ (\ref{eq:Cbgap}).
On the other hand, if we first take the limit $ \omega \rightarrow 0$ and then
consider the leading behavior at low temperatures, we obtain for
$ T \ll \Omega_2$
 \begin{equation}
 \nu ( 0, T  ) \sim 2  C \frac{T }{ e^4} 
 \; ,
 \label{eq:nuasymT}
 \end{equation}
where the numerical constant $C$ is given in Eq.\ (\ref{eq:Cres}).
Note that the prefactor in Eq.\ (\ref{eq:nuasymT}) is twice as large
as the corresponding prefactor in Eq.\ (\ref{eq:nuasym}).
Thus, at low temperatures the average DOS at the Fermi energy 
vanishes linearly in $T$, which resembles the behavior of the average DOS
in the localized regime.
A plot of the DOS for various temperatures is shown in Fig.\ \ref{fig:2d:DOS}. 

\begin{figure}[htb]
\begin{center}
\psfrag{T}{\hspace{1.5mm}$T/\Omega_2$}
\psfrag{dos}{$\nu(T )/\nu_0$}
\psfrag{0}{$0$}
\psfrag{0.2}{$0.2$}
\psfrag{0.1}{$0.1$}
\psfrag{0.3}{$0.3$}
\psfrag{0.4}{$0.4$}
\psfrag{0.5}{$0.5$}
\psfrag{0.6}{$0.6$}
\psfrag{0.8}{$0.8$}
\psfrag{1}{$1$}
\epsfxsize8.0cm
\epsfbox{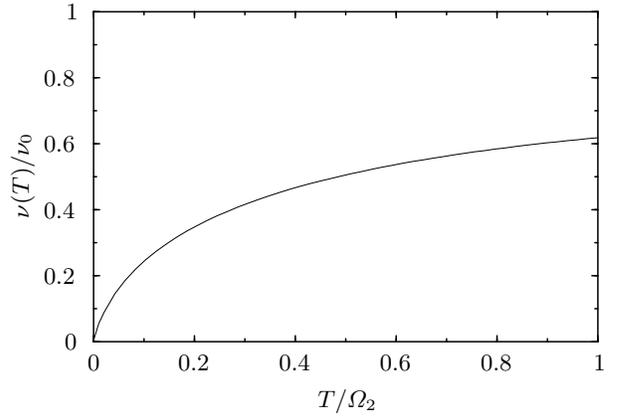}
\caption{Temperature-dependence of the average DOS  $\nu(T)$ 
at the Fermi energy in $2d$.  
The parameters are again
$r_0 \to r_{\ast} =1/\pi$ and $\tau_1 = \tau_0$. }
\label{fig:2d:DOST}
\end{center}
\end{figure}

Formally,   the derivation of Eq.\ (\ref{eq:2d:DOS})  is only valid in the limit
of weak disorder where $r_0  = 1/\pi k_F \ell  \ll 1$. However,
if the conductivity has a finite DC limit
$ \sigma ( 0 )$, then it is reasonable to expect  that the qualitative behavior of the DOS
can be obtained by simply replacing
$r_0 \rightarrow r_{\ast}  = (e^2/h)/\pi \sigma (0)$ in Eq.\ (\ref{eq:2d:DOS}).
In Ref. \cite{Kopietz98}
this replacement has been justified via a simple renormalization group argument.
For a possible comparison of our results with
future tunneling experiments on semiconductor materials which apparently
show a metal-insulator transition \cite{Kravchenko95,Abrahams01}
(as far as we know, such experiments have not been performed yet),
 we have plotted  in Fig.\ \ref{fig:2d:DOS} the frequency-dependence
predicted by Eq.\ (\ref{eq:2d:DOS})
for $r_0 \rightarrow r_{\ast} = 1/ \pi $, corresponding
to the zero-temperature conductivity of order $\sigma (0) \approx e^2/h$.
Saturation values in this regime are typically encountered 
in the metallic regime close to  the apparent metal-insulator transition \cite{Abrahams01}.
The temperature-dependence of the average DOS at the Fermi energy
is shown in Fig.\ \ref{fig:2d:DOST}.

It should be noted that in deriving Eq.\ (\ref{eq:2d:DOS}) we have
consistently neglected corrections involving single logarithms. 
We cannot exclude the possibility that these corrections
 lead to a further depletion of the Coulomb gap. For example, 
taking the (sub-leading) imaginary part of the full Debye-Waller factor given 
in Eq.\ (\ref{eq:d2:DebyeWaller2}) into account would lead 
to a nonlinear suppression of the DOS. However, as discussed above, such an
approximation would not be systematic.

\section{Zero bias anomaly in 1d}
\label{sec:1d}

As discussed in Sec.\ \ref{sec:gauge}, the naive application of
Eqs.\ (\ref{eq:G=G0*exp}-\ref{eq:DebyeWaller}) in quasi $1d$ 
is problematic, because in this case the interaction corrections to the DOS and to the 
conductivity both involve the same type of singularities.
Keeping this caveat in mind, let us 
nevertheless briefly discuss the predictions of
Eqs.\ (\ref{eq:G=G0*exp}-\ref{eq:DebyeWaller}) in quasi $1d$.
A similar calculation has recently been performed in Ref. \cite{Mishchenko01}.

\begin{figure}
\begin{center}
\psfrag{omega}{$\omega/\Omega_{1}$}
\psfrag{dos}{$\nu(\omega)/\nu_0$}
\psfrag{0}{$0$}
\psfrag{0.2}{$0.2$}
\psfrag{0.4}{$0.4$}
\psfrag{0.6}{$0.6$}
\psfrag{0.8}{$0.8$}
\psfrag{1}{$1$}
\psfrag{2}{$2$}
\psfrag{3}{$3$}
\psfrag{4}{$4$}
\psfrag{5}{$5$}
\psfrag{6}{$6$}
\psfrag{8}{$8$}
\psfrag{10}{$10$}
\psfrag{15}{$15$}
\psfrag{20}{$20$}
\psfrag{-10}{$-10$}
\psfrag{-20}{$-20$}
\epsfxsize8.0cm
\epsfbox{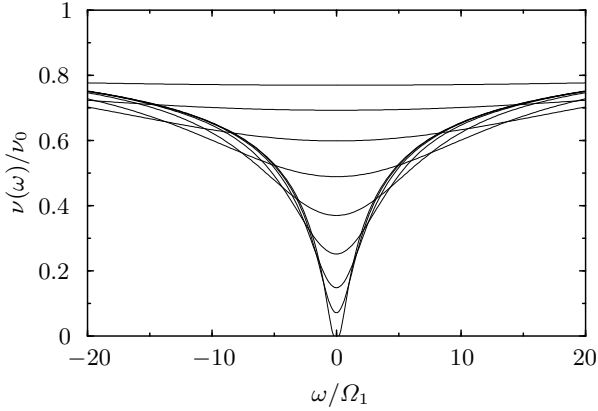}

\caption{Graph of the average DOS $\nu(\omega)$ in $1d$ 
for various temperatures, see
Eq.\ (\ref{eq:DOS}). The curves from 
top to bottom are for $T/ \Omega_1 = 32,16,8,4,2,1,0.5,0.25,0$.}
\label{DOS:1d}
\end{center}
\end{figure}

\begin{figure}
\begin{center}
\psfrag{T}{\hspace{1.8mm}$T/\Omega_1$} 
\psfrag{dos}{$\nu(T)/\nu_0$}
\psfrag{0}{$0$}
\psfrag{0.2}{$0.2$}
\psfrag{0.4}{$0.4$}
\psfrag{0.6}{$0.6$}
\psfrag{0.8}{$0.8$}
\psfrag{1}{$1$}
\psfrag{2}{$2$}
\psfrag{3}{$3$}
\psfrag{4}{$4$}
\psfrag{5}{$5$}
\psfrag{6}{$6$}
\psfrag{8}{$8$}
\psfrag{10}{$10$}
\psfrag{15}{$15$}
\psfrag{20}{$20$}
\epsfxsize8.0cm
\epsfbox{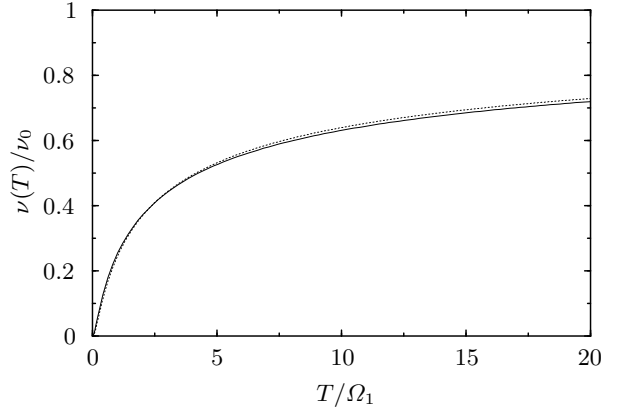}
\caption{Solid line: graph of the average DOS $\nu(T)$ for $\omega =0$, see Eq.\ (\ref{eq:DOS}). 
Dotted dotted line: the approximation $\exp(-\sqrt{2\Omega_1/T})$.}
\label{fig2}
\end{center}
\end{figure}

Since screening is much less effective in one dimension than in 
higher dimensions, we simply approximate the
one-dimensional RPA-interaction by a constant,
$f^{\rm {RPA}}_{q,\omega_m} \approx f_0$ \cite{Altshuler85}. 
This approximation should be correct up to logarithmic corrections in frequency.
A  calculation analogous to the one leading  to Eq.\ (\ref{eq:d2:DebyeWaller2}) results in
\begin{equation}
  \label{eq:DebyeWaller2}
  Q(it) = -  \sqrt{2 \Omega_1 }  \left[ 
 \int_0^{\infty} \frac{d\omega}{\sqrt{2\pi}} \,\frac{1-\cos(\omega t)}{\omega^{3/2} \tanh(\beta \omega/2)} + i \sqrt{ t} \right] \;,
\end{equation}
where $\Omega_1$ is given in Eq.\ (\ref{eq:omega.ast}).
Note that in $1d$, there is no need for an ultraviolet cutoff. At $T=0$, we have $\textrm{Re}\, Q(it) = \textrm{Im}\, Q(it)$, such that the imaginary part of $Q(it)$ cannot be neglected. The DOS at finite temperature $T = 1/\beta$ can now be written as
\begin{eqnarray}
  \label{eq:DOS}
  \nu(\omega, T ) 
& \approx & \nu_0 \coth \left( \frac{\beta \omega}{2} \right) 
\frac{2}{\beta} \int_0^{\infty} dt \, \frac{\sin (\omega t)   
\cos (\sqrt{2 \Omega_{1} t} )   }{\sinh(\pi t/ \beta)} 
\nonumber \\
& & \hspace{-10mm} \times 
\exp{\left[ -\sqrt{  \frac{ \Omega_{1}}{\pi} } \int_0^{\infty} 
d \omega^{\prime}  \,
\frac{1-\cos( \omega^{\prime} t)}{ ( \omega^{ \prime })^{3/2}  
\tanh(\beta  \omega^{\prime} /2)} 
\right]} 
 \; .
\end{eqnarray}
As can be easily checked, for $|\omega| \gg {\rm max} \{ \Omega_1,T\}$, 
Eq.\ (\ref{eq:DOS}) reduces to the perturbative result given in Eq.\ (\ref{eq:Altshuler1}). 
A graph of $\nu(\omega, T )$ 
for different  temperatures is shown in Fig.\ \ref{DOS:1d}. 
The DOS at the Fermi energy as a function of temperature
is shown in Fig.\ \ref{fig2}.
The DOS  is well approximated by 
$\nu(T) \approx \nu_0 \exp(-\sqrt{2\Omega_{1}/T})$ which is quite similar to a result found in Ref.\ \cite{Mishchenko01}.
At zero temperature, Eq.\ (\ref{eq:DOS}) may be simplified to
\begin{equation}
  \label{eq:DOS.T=0}
  \nu(\omega, T = 0) = \frac{\nu_0}{\pi}  \textrm{Re}\ \int_0^{\infty} dx\, \frac{\sin x^2}{x}
 \exp\left[-2 x \sqrt{\frac{i \Omega_{1}}{|\omega|}} \ \right] \;.
\end{equation}
For $|\omega| \ll \Omega_1$, 
this integral is easily evaluated within a saddle-point approximation,
\begin{equation}
  \label{eq:DOS.T=0.smallomega}
  \nu(\omega, T=0) \sim \nu_0 \sqrt{\frac{|\omega|}{\pi \Omega_{1}}} \, 
\exp \left[-\frac{\Omega_{1}}{|\omega|} \right]\; ,\; \;  |\omega| \ll \Omega_{1} \;,
\end{equation}
which differs from the corresponding expression given by 
Mishchenko {\em et al.}  \cite{Mishchenko01} (see Eq.\ (\ref{eq:numishchenko}))
by a different prefactor.
The exponential suppression of the DOS is quite surprising and is possibly an artifact
of the inconsistent exponentiation of the
$ | \omega |^{-1/2}$-singularities inherent in
Eqs.\ (\ref{eq:DOS.T=0.smallomega}) and (\ref{eq:numishchenko}) 
for quasi $1d$ systems.
Recall that for long-range Coulomb interactions  in $2d$
it is possible to separate $\ln^2$-terms from sub-leading terms already 
neglected in the approximation made in Eq.\ (\ref{eq:G=G0*exp}).
In $1d$, a similar separation is not possible.
 The ultimate low-frequency behavior of the DOS in $1d$ 
might therefore be altered by diagrams neglected in Eq.\ (\ref{eq:G=G0*exp}).

\section{Summary and conclusions}

In summary, we have considered the fate  of the
singular perturbative corrections to the average DOS in
quasi $1d$ and $2d$ disordered metals  when the frequency and the temperature
are reduced such that one leaves the perturbative regime.
For a $2d$ metal with long-range Coulomb interactions, it is possible to re-sum the
leading $\ln^2$-singularities in the perturbative expansion of the average DOS 
to all orders in perturbation theory. The method relies on the fact that similar singularities do not
appear in the perturbative expansion of the average conductivity.
For a $2d$ system with short-range interactions or for a quasi $1d$ system 
a similar separation of energy scales does not exist, so that 
approximate expressions of the type given in Eq.\ (\ref{eq:G=G0*exp}), which
involve the exponentiation of a certain subclass of Feynman diagrams, 
are problematic in this case.
Thus, in practice the gauge-transformation trick described above, which has
recently been 
employed by many authors \cite{Levitov97,Kopietz98,Kamenev99,Nazarov89},
is only controlled in the case of $2d$ disordered electrons
interacting with {\it{long-range}} Coulomb forces in the regime where the
conductivity has a finite DC-limit.
In this case this method yields
a simple interpolation formula (\ref{eq:2d:DOS}) for the average DOS,
which predicts a smooth crossover from the perturbative regime at high frequencies
to a  new low-frequency regime, where $\nu ( \omega , T )$
vanishes linearly in $\omega $ or $T$.
The average DOS of {\it{localized}} classical electrons is known to show 
a similar frequency- or temperature-dependence \cite{Efros85}.
In contrast,  the metallic Coulomb gap discussed here has a quantum mechanical origin and requires delocalized electrons
with a finite DC conductivity.  Numerical evidence for such a
metallic Coulomb gap has been found in Ref. \cite{Efros95}.
 
In principle, it should be possible 
to verify the existence of the metallic Coulomb gap experimentally via tunneling experiments
in strongly correlated disordered systems with a finite DC conductivity.
The expected shape of a typical trace of the tunneling conductance
as a function of the applied voltage  is shown in Fig.\ \ref{fig:2d:DOS}.
Recent tunneling experiments by Bielejec {\it{et al.}} \cite{Bielejec01}
  in quench-condensed Beryllium films
show a crossover from the perturbative regime with logarithmic corrections 
to an apparently
linear Coulomb gap in the DOS. 
However, at the lowest temperatures the DOS exhibits a hard correlation gap,
the origin of which remains open.



\begin{thebibliography}{99}
%
\bibitem{Altshuler85} B.\ L.\ Altshuler and A.\ G.\ Aronov, in {\em Electron-electron interactions in disordered systems}, edited by A.\ L.\ Efros and M.\ Pollak (North-Holland, Amsterdam, 1985).
%
\bibitem{Bielejec01}
E.\ Bielejec, J.\ Ruan, and W.\ Wu, Phys.\ Rev.\ Lett.\ {\bf 87}, 036801 (2001). 
%
\bibitem{Bachtold99} A.\ Bachtold, C.\ Strunk, J.\ P.\ Salvetat, J.\ M.\ Bonard, L.\ Forr{\'o}, T.\ Nussbaumer, and C.\ Sch{\"o}nenberger, Nature {\bf 397}, 673 (1999); C.\ Sch{\"o}nenberger, A.\ Bachtold, C.\ Strunk, J.\ P.\ Salvetat, and L.\ Forr{\'o}, Appl.\ Phys.\ A {\bf 69}, 283 (1999).
%
\bibitem{Bachtold00} A.\  Bachtold, M.\  Fuhrer, S.\  Plyasunov, M.\  Forero, E.\ H.\  Anderson, A.\  Zettl, and P.\ L.\  McEuen, Phys.\  Rev.\  Lett.\  {\bf 84}, 6082 (2000).
%
\bibitem{Kravchenko95}
S.\ V.\ Kravchenko, W.\ E.\ Mason, G.\ E.\ Bowker, and J.\ E.\ Furneaux,
Phys.\ Rev.\ B {\bf{51}}, 7038 (1995).
%
\bibitem{Abrahams01}
For a recent review see 
E.\ Abrahams, S.\ V.\ Kravchenko, and M.\ P.\ Sarachik, Rev.\ Mod.\ Phys.\ {\bf 73}, 251 (2001).
%
\bibitem{Finkelstein83}
A.\ M.\ Finkelstein, Zh.\ Eksp.\ Teor.\ Fiz.\ {\bf{84}}, 168 (1983)
[Sov.\ Phys.\ JETP {\bf{57}}, 97 (1983)].
%
\bibitem{Belitz93}
D.\ Belitz and T.\ R.\ Kirkpatrick, Phys.\ Rev.\ B {\bf{48}}, 14072 (1993).
%
\bibitem{Levitov97}
S.\ Levitov and A.\ V.\ Shytov, Prisma Zh.\ Eksp.\ Teor.\ Fiz.\ {\bf 66}, 200 (1997) [JETP Lett.\ {\bf 66}, 214 (1997)].
%
\bibitem{Kopietz98} 
P.\ Kopietz, Phys.\ Rev.\ Lett.\ {\bf{81}}, 2120 (1998).
%
\bibitem{Kamenev99}
A.\ Kamenev and A.\ Andreev, Phys.\ Rev.\ B {\bf{60}}, 2218 (1999).
%
\bibitem{Efros85}
A.\ L.\ Efros and B.\ I.\ Shklovskii,
in {\it{Electron-Electron Interactions in Disordered Systems}},
edited by A.\ L.\ Efros and M.\ Pollak (North-Holland, Amsterdam, 1985).
%
\bibitem{Efros95}
A.\ L.\ Efros and F.\ G.\ Pikus, Solid State Commun.\ {\bf{96}}, 183
(1995).
%
\bibitem{Benenti99}
G.\ Benenti, X.\ Waintal, and J.-L.\ Pichard, Phys.\ Rev.\ Lett.\ {\bf 83}, 1826 (1999); 
X.\ Waintal, G.\ Benenti, and J.-L.\ Pichard, Europhys.\ Lett.\ {\bf 49}, 466 (2000);
G.\ Benenti, X.\ Waintal, J.-L.\ Pichard, and D.\ L.\ Shepelyansky, Eur.\ Phys.\ J.\ B {\bf 17}, 515 (2000).
\bibitem{Nazarov89}
Y.\ V.\ Nazarov, Zh.\ Eksp.\ Teor.\ Fiz.\ {\bf{96}}, 975 (1989)
[Sov.\ Phys.\ JETP {\bf{68}}, 561 (1989)].
%
\bibitem{Mishchenko01}
E.\ G.\ Mishchenko, A.\ V.\ Andreev, and L.\ I.\ Glazman, cond-mat/0106448.
%
\bibitem{Egger01}
R.\ Egger and A.\ O.\ Gogolin, Phys.\ Rev.\ Lett.\ {\bf 87}, 066401 (2001).
%
\bibitem{footnote1}
Actually, in Ref.\ \cite{Kopietz98} the factor $D_0 {\bf{q}}^2$ in the denominator
of Eq.\ (\ref{eq:DebyeWaller}) does not explicitly appear; instead, 
the frequency-summation is restricted to the regime
$| \omega_m | / D_0 \kappa \ll | {\bf{q}} |  \ll ( | \omega_m | / D_0 )^{1/2}$.
Both prescriptions lead to identical results as far as
the summation of the leading $\ln^2$-singularities are concerned.
Here we implement the cutoff as in Eq.\ (\ref{eq:DebyeWaller}), because
then  Eqs.\ (\ref{eq:G=G0*exp}) and  (\ref{eq:DebyeWaller}) are identical 
with the  result obtained by Kamenev and Andreev within a field theoretic
approach based on the Keldysh formalism \cite{Kamenev99}.


\end{thebibliography}
\end{document}